\documentclass[aps,epsf,twocolumn,floatfix,showpacs,prl]{revtex4}

\newcommand{\ket}[1]{\left |#1\right >}

\newcommand{\expec}[1]{\left < #1\right >}

\usepackage{amsmath}
\usepackage{amsfonts}
\usepackage{graphicx}
\usepackage{amsthm}
\usepackage{subfigure}

\usepackage{hyperref}

\renewcommand{\v}[1]{\boldsymbol{#1}}

\newcommand{\lp}{\left ( }
\newcommand{\rp}{\right ) }

\newcommand{\hc}{\text{H.c.}}

\newcommand{\beq}{\begin{eqnarray*}}
\newcommand{\eeq}{\end{eqnarray*}}

\newcommand{\be}{\begin{eqnarray}}
\newcommand{\ee}{\end{eqnarray}}

\def\lsim{\mathrel{\rlap{\lower4pt\hbox{\hskip1pt$\sim$}}
    \raise1pt\hbox{$<$}}}                

\def\gsim{\mathrel{\rlap{\lower4pt\hbox{\hskip1pt$\sim$}}
    \raise1pt\hbox{$>$}}}                

\begin{document}


\bibliographystyle{prsty}

\title{Explanation of 100-fold reduction of spectral shifts for hydrogen on helium films}
\author{Kaden R.A. Hazzard} \email{kh279@cornell.edu}
\affiliation{Laboratory of Atomic
and Solid State Physics, Cornell University, Ithaca, New York 14853, USA}

\author{Erich J. Mueller}
\affiliation{Laboratory of Atomic and Solid State Physics, Cornell
University, Ithaca, New York 14853, USA}

\begin{abstract}

We show that helium film-mediated hydrogen-hydrogen interactions account for a two orders of magnitude discrepancy between previous theory and recent experiments on cold collision shifts in spin-polarized hydrogen adsorbed on a helium film. These attractive interactions also explain the anomalous dependence of the cold collision frequency shifts on the $^3$He covering of the film.  Our findings suggest that the gas will become mechanically unstable before reaching the Kosterlitz-Thouless transition unless the experiment is performed in a drastically different regime, for example with a much different helium film geometry.

\end{abstract}

\maketitle

Two-dimensional (2d) spin polarized hydrogen is an exciting quantum fluid.  
It can support a
superfluid-normal Kosterlitz-Thouless transition, which crosses over to 
Bose-Einstein condensation  as the third-dimensional trapping is varied. Hydrogen experiments are unique in involving techniques from both low-temperature physics and atomic physics.  The large particle numbers and light mass give a  
high Bose-Einstein condensation temperature,  and the interaction properties are known to high accuracy.

In this context, it is remarkable that experiments on spin-polarized hydrogen adsorbed on a helium film display a cold collision frequency shift which is two orders of magnitude smaller than theory predicts~\cite{vasiliev:h2_on_he_expt}.  Here we show that a helium mediated interaction can explain this discrepancy and a similarly unexpected dependence of the shift on the concentration of $^3$He in the film.

Since these experiments
   are in the ``cold collision" regime~\cite{pethick:pseudopot-breakdown}, where the  thermal de Broglie wavelength is longer than the interaction's effective range, they see extremely sharp spectral lines which are shifted by interactions.
These shifts are important: they limit 
hydrogen maser~\cite{walsworth:masers,hayden:masers} and atomic fountain~\cite{gibble:fountain} performance; the 1S-2S spectral shift
first demonstrated Bose-Einstein condensation in hydrogen~\cite{fried:hydrogen}; the Mott insulator-superfluid phase transition was probed in atomic gases in optical lattices~\cite{hazzard:clock-shift-bh,campbell:ketterle-clock-shift}; and the shifts revealed features of two-component Fermi gases across a Feshbach resonance~\cite{schunck:pairing-no-sf,shin:tomographic-rf}.  Such spectra are promising probes of future strongly-correlated atomic systems.

\textit{Calculating spectra.}---Letting $\ket{ij}$ be the state with electron spin $i$ and nuclear spin $j$,
the relevant hydrogen spin states in the experimental $B=4.6$T field are $\ket{1}\equiv\cos \theta\ket{\downarrow\uparrow}-\sin\theta\ket{\uparrow\downarrow}$  and $\ket{2}\equiv\cos\theta\ket{\uparrow\downarrow} +\sin\theta\ket{\downarrow\uparrow}$ with $\theta$ negligible.  Initially all of the atoms are in state $|1\rangle$.
Neglecting, for now, the helium film's degrees of freedom, the quasi-2d hydrogen's Hamiltonian is
\begin{equation}
\!\!H\! =\! \sum_{j,\v{k}} \epsilon_{j,\v{k}} \psi^\dagger_{j,\v{k}} \psi_{j,\v{k}} 
+  
\!\! 
\sum_{\buildrel {i,  j}\over {\v{k},\v{p},\v{q} }}   \!\!  \frac{V_{ij,\v{q}}}{2} \, \psi_{j,\v{p}}^\dagger \psi_{i,\v{k}}^\dagger\psi_{i,\v{k}+\v{q}}\psi_{j,\v{p}-\v{q}} \label{eq:hydrogen-ham}
\end{equation}
with $\psi^\dagger_{j,\v{k}}$ bosonic creation operators for states of momentum $\v{k}$ and internal state $j$;  $\epsilon_{j,\v{k}}=
 k^2/2m+\delta_j-\mu_j$ the free dispersion of the effectively 2d gas, including the internal energy $\delta_j$ and chemical potential $\mu_j$; and $V_{ij,\v{k}}$ the Fourier space interaction potential between atoms in states $i$ and $j$. Throughout we set $\hbar=1$.

The spectrum is measured by counting the number of atoms transferred from state $|1\rangle$ to state $|2\rangle$ after applying a frequency $\omega$ electromagnetic probe for a short time.  In Ref.~\cite{vasiliev:h2_on_he_expt}, a combination of electron spin resonance and nuclear magnetic resonance  drives the transition. In the rotating wave approximation the probe beam's Hamiltonian~\cite{fn:chem-pot} is
$
H_P = \Omega_P \sum_{\v{k}} e^{-i(\omega-(\mu_2-\mu_1))t}\psi_{2,\v{k}}^\dagger \psi_{1,\v{k}} + \hc
$
The radio frequency and microwave photons transfer negligible momentum.

Given that the range of the the potential ($\sim\!1\AA$) is significantly less than the 2d interparticle separation ($\sim\!100\AA$) the average absorbed photon energy shift is given in terms of the scattering amplitude~\cite{oktel:cs-ref,oktel:cs-ref2}
\be
\delta \omega &=& \frac{1}{m}g_2(0)\lp f_{12}-f_{11}\rp \sigma_1, \label{eq:coherent-clock-shift-fundamental-3d} 
\ee
with $g_2(r)\equiv\expec{\psi_1^\dagger(r)\psi_1^\dagger(0)\psi_1(0)\psi_1(r)}/
\expec{\psi_1^\dagger(0)\psi_1(0)}^2$, $f_{ij}$ the $i$-$j$ scattering amplitude evaluated at characteristic momenta of the system (here, set by the thermal de Broglie momentum), and $\sigma_1\equiv\expec{\psi_1^\dagger(0) \psi_1(0)}$  the 2d density.  In a dilute thermal gas $g_2(0)=2$. A similar formula holds in 3d, where the scattering amplitude is $f_{ij}=4\pi a_{ij}$ with $a_{ij}$ the 3d scattering length. In the case we consider, the spectrum is a delta function~\cite{pethick:pseudopot-breakdown}.

In quasi-2d, where  kinematics are 2d but the 3d scattering length is much less than the perpendicular confinement length, we can construct 
 a 2d interaction giving identical  
 low-energy 
scattering properties.  Assuming harmonic confinement with oscillation frequency $\omega_{\text{osc}}$, the effective 2d
scattering amplitude is~\cite{petrov:2d_scaling,petrov:2d_scaling_short,petrov:2d_scaling_full}
\be
f &=& 2\sqrt{2\pi}\frac{1}{l/a - (1/\sqrt{2\pi})\log (\pi q_T^2 l^2)}\label{eq:2d-coupling}
\ee
where $q_T \equiv \sqrt{m k_B T}
$ is the thermal momentum and $l\equiv \sqrt{\hbar /(m\omega_{\text{osc}})}$ is the length scale of $z$-axis confinement of the hydrogen gas.
  Following Ref.~\cite{vasiliev:h2_on_he_expt}'s discussion, the characteristic length scale for the confinement in their experiments
is $l_0=\sqrt{2\pi}\ell=
1/\sqrt{2mE_a}\sim 5\AA$ where $E_a$ is the adsorption energy of the hydrogen on the helium film.  
The spectral shift is then
\be
\delta \omega
    &=&  \frac{4\pi
    }{m}\bigg[\frac{1}{l_0/a_{12} - \log (q_T^2 l_0^2/2)}\nonumber\\
    &&\hspace{0.4in}{}-\frac{1}{l_0/a_{11} - \log (q_T^2 l_0^2/2)} \bigg] g_2(0) \sigma_1.\label{eq:coherent-clock-shift-fundamental-2d-no-simple}
\ee
If
$a\ll l_0$
this reduces to the simpler form
\be
\delta \omega &=& \frac{4\pi
}{m}g_2(0)\lp a_{12}-a_{11}\rp\frac{\sigma_1}{l_0},\label{eq:coherent-clock-shift-fundamental-2d}
\ee
as for a 3d gas with density $\sigma_1/l_0$.

The spectral shift in Eq.~(\ref{eq:coherent-clock-shift-fundamental-2d}) is one hundred times larger than experiment.  
We will show that this discrepancy is consistent with the corrections found by including the hydrogen-hydrogen interaction mediated by the helium film, $V_{\text{med}}$.  Although $V_{\text{med}}$ is state-independent, it
alters $\delta \omega$ by reducing the probability that two particles are close enough to feel the spin dependent $V_{ij}$.


\textit{Helium film-mediated interaction.}---Reference~\cite{wilson:surface_induced_interaction} derived  a hydrogen-only effective action assuming that (i) the helium film's excitations are non-interacting, (ii) the hydrogen-helium interaction falls off as $1/r^6$, and (iii) that the hydrogen confinement length $l_0$ is significantly smaller than the hydrogen-helium separation $\zeta$.
 Defining the total hydrogen density operator $\rho_t(\v{\rho})\equiv \sum_j \psi_j^\dagger(\v{\rho})\psi_j(\v{\rho})$, the Fourier transform of Ref.~\cite{wilson:surface_induced_interaction}'s effective Hamiltonian is
 \begin{equation}
H_{\text{eff}}
=  H
-\frac{1}{2} \int \!\!\int d^2 \! \rho\,d^2\!\rho'\,\rho_{\text{t}}(\v{\rho}) \rho_{\text{t}}(\v{\rho'}) V_{\text{med}}(\v{\rho}-\v{\rho'})\label{eq:med-pot-c}
\end{equation}
with mediated pair interaction given by
\be
V_{\text{med}} &=& V_0 {\bar V}_{\lambda/\zeta}(\v{R}/\lambda)
\ee
in terms of 
\be
V_0 &\equiv& \frac{2\delta^2}{\pi^3 \lambda^2 MC_3^2}\label{eq:V0-defn}
 \ee
 with $\Lambda_0$ defined as the parameter controlling the hydrogen-helium van der Waals interaction strength such that the potential is $V_{\text{H-He}}=-(6\Lambda_0/n\pi) r^{-6}$ with $n$ the helium density, and with
$\delta \equiv (6\Lambda_0 \phi_g/n\zeta^4)$,
 $\lambda\equiv \sqrt{\frac{1}{MC_3^2}}\lp \frac{1}{2M}+\frac{\beta d_0}{n}\rp$, $\zeta$ the distance from the hydrogen gas to the helium film, $C_3$ the film's third sound speed, $M$ the helium mass, $d_0$ the film thickness,  $\beta$ the film surface tension, and $\phi_g\equiv\sqrt{n d_0}$  (estimates of parameters in experiments are given later).  The non-dimensionalized potential $V_\xi$ is defined as
\be
\hspace{-0.2in} \bar V_\xi(x) \hspace{-0.03in}&=& \hspace{-0.03in}\int\!\!\int d^2\bar \rho \,d^2\bar \rho'\, \bar A(\bar\rho) \bar A(\bar \rho') \bar G_F\lp \frac{\v{x}+\v{\bar \rho} + \v{\bar \rho'}}{\xi}\rp,\label{eq:mediated-potential-rescaled-b}
\ee
where we have defined
\be
{\bar A}(x) &\equiv& \frac{1}{\lp 1+x^2\rp^3} \hspace{0.15in}\text{ and } \hspace{0.15in}{\bar G_F}(x) \equiv K_0(x),
\ee
with $K_0$ the zero'th modified Bessel function.  Note that $A(\rho)=-(12\phi_g \Lambda_0/\pi n \zeta^6){\bar A}(\rho/\zeta)$ is the hydrogen-helium van der Waals interaction at a separation $\rho$ and $G_F(\rho)=(n/2d_0 \pi \beta) {\bar G_F}(\rho/\lambda)$ is the helium film's Green's function.

Eq.~(\ref{eq:med-pot-c}) neglects $V_{\text{med}}$'s frequency dependence,
 which arises from retardation: one hydrogen atom excites a helium excitation which  propagates to another hydrogen atom.  In a future work we will explore retardation effects, but here we neglect them.

\begin{figure}[hbtp]
\setlength{\unitlength}{1.0in}
\centering
\subfigure[]{
\includegraphics[width=1.62in,angle=0]{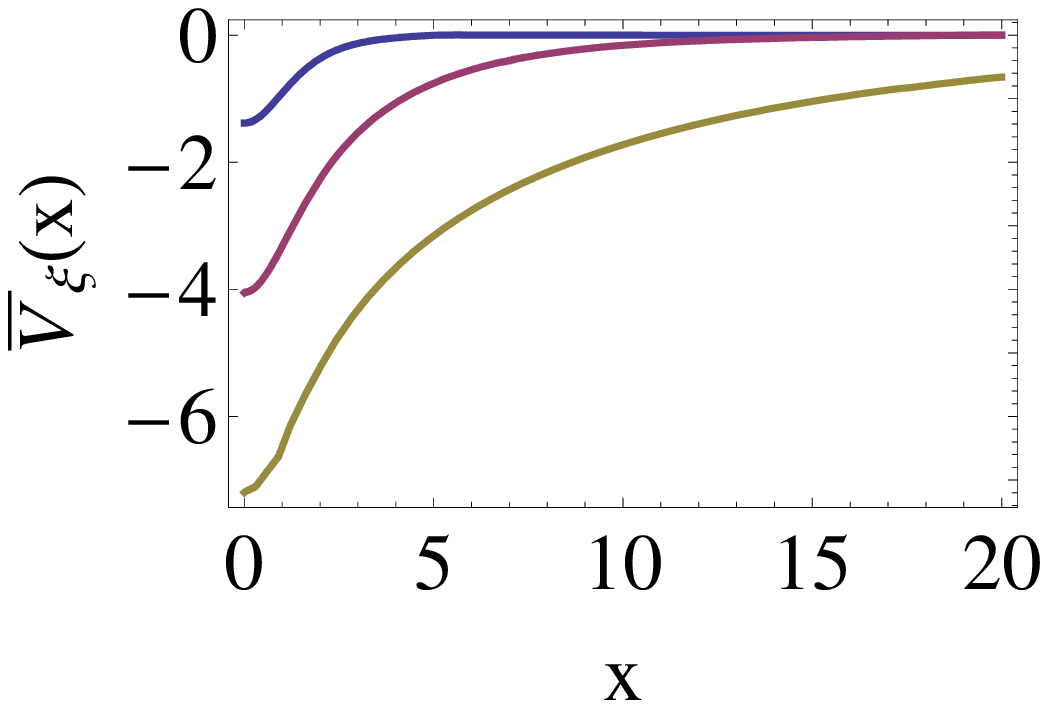}\label{fig:typical-pot}
}\hspace{-0.07in}
\subfigure[]{
\includegraphics[width=1.62in,angle=0]{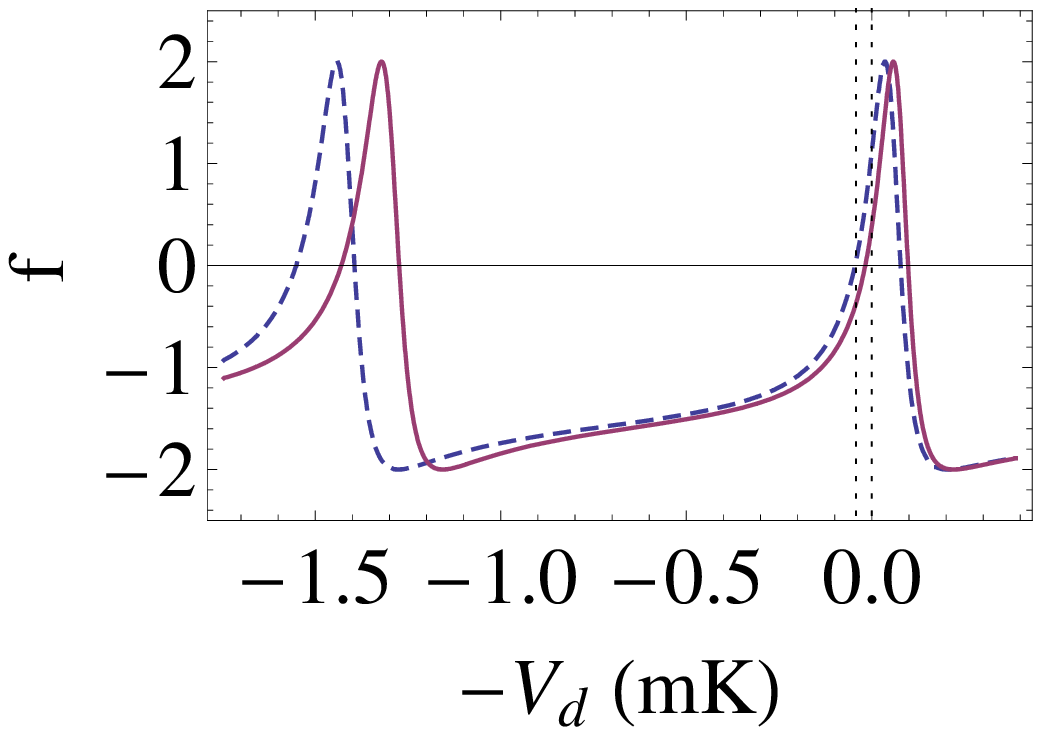}\label{fig:coupling-constant-plot}
}\vspace{0.05in}
\subfigure[]{
\put(-1.45,0){\includegraphics[width=3.0in,angle=0]{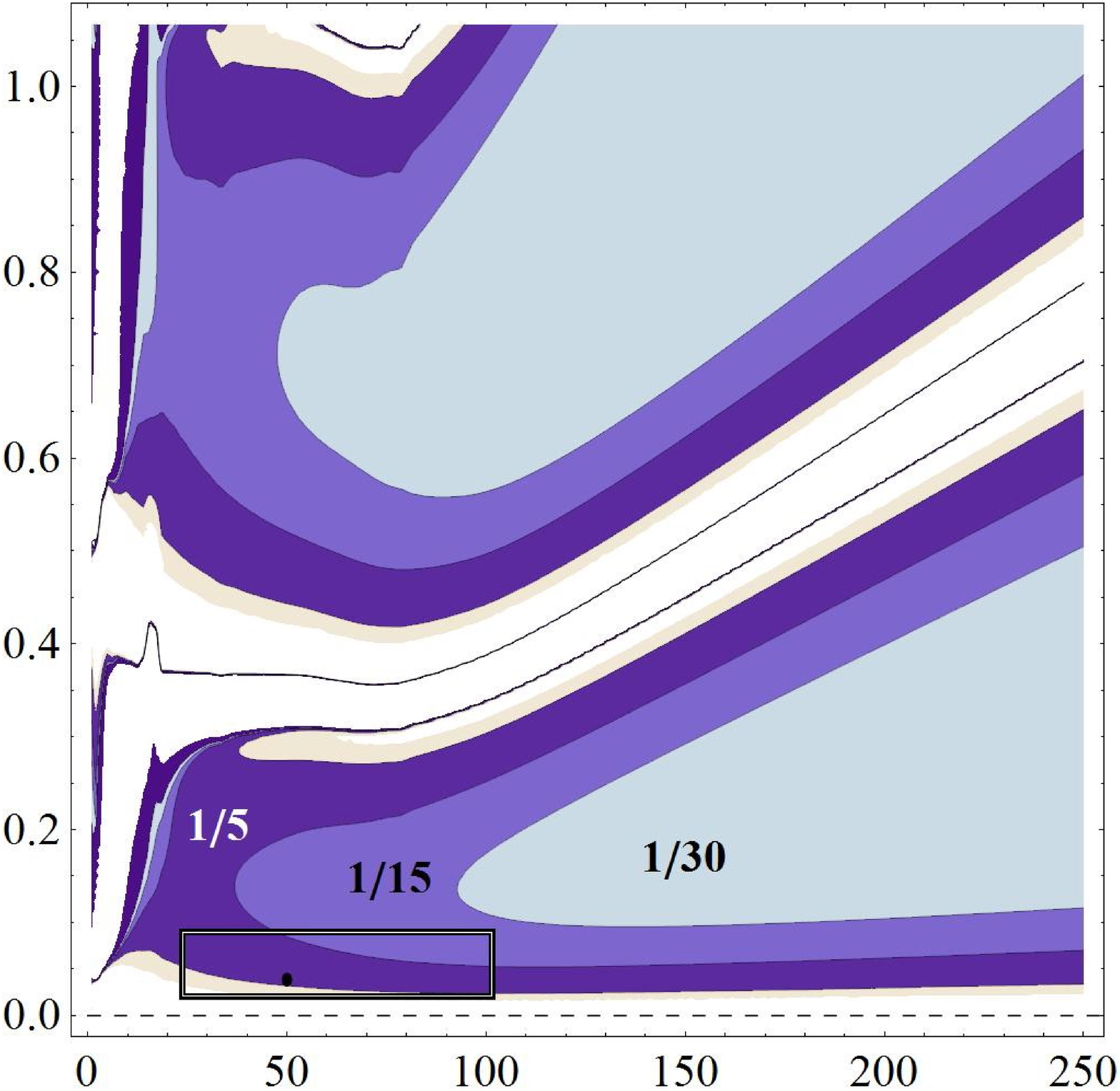}\label{fig:lambda-strength-coupling-plot}
}
\put(-1.7,1.25){\rotatebox{90}{$V_d$ (mK)}}
\put(0.09,-0.13){$\lambda$}
}
\caption{ (a) Rescaled mediated potential as a function, as given by Eq.~(\ref{eq:mediated-potential-rescaled-b}), with $\xi=1,4,15$ from top to bottom. (b) Scattering amplitudes $f$ as a function of mediated potential depth $V_d \equiv (2\delta^2)/(\pi^3 \lambda^2 MC_3^2)$ for the triplet (dashed) and singlet (solid) scattering channels. Vertical lines indicate (i) the strength  of the mediated potential using the typical parameters given in the text and (ii) the absence of a mediated potential.  The horizontal axis's units are converted assuming $\lambda=50\AA$, with $\lambda$ the helium film Green function's characteristic length.
(c) Frequency shift reduction as a function of potential depth scale factor $V_d =(2\delta^2)/(\pi^3 \lambda_0^2 MC_3^2)$, with $\lambda_0\equiv 50\AA$ and characteristic length $\lambda$.  Typical values of parameters are shown by the  red dot, while factor of two variations comprise the interior of the red rectangle.  Contour lines correspond to reductions of factors of $1/5$, $1/15$, and $1/30$.
}
\end{figure}

We numerically compute $\bar V_\xi(x)$ as a function of $x$ and $\xi$.
Typical results are shown in Fig.~\ref{fig:typical-pot}.

\textit{Spectral line shifts with mediated potential.}---To evaluate spectral shifts via Eq.~(\ref{eq:coherent-clock-shift-fundamental-3d}), we calculate the scattering amplitude of  $V_{\text{tot}}\!=\!V_{ij}+V_{\text{med}}$.  
Since the bare hydrogen-hydrogen potential is short-ranged compared to $V_{\text{med}}$ and the depth of  $V_{\text{med}}$ will be sufficiently low, we may replace $V_{\text{tot}}$ with $V_{\text{tot}}'\!=\!V_{ij}'+V_{\text{med}}$, where $V_{ij}'$  is an arbitrary short range potential reproducing $V_{ij}$'s scattering amplitude.
We numerically solve the resulting two-particle Schr\"odinger equation with potential  $V_{\text{tot}}$, replacing the  potential $V_{ij}$ with a  boundary condition at small interparticle separations.

We use the following estimates in our calculations, taken from Ref.'s~\cite{eijnde:surface-relaxation,wilson:surface_induced_interaction,vasiliev:h2_on_he_expt}:
$l_0\!\sim \!5\AA$, $a_t=0.72 \AA$, $a_s=0.17\AA$, $\lambda\!\sim\!50\AA$, $C_3\!\sim\! 1\,$m/s $q_T=\sqrt{m k_B T}
\!\sim \!(30\AA)^{-1}$, $ \delta \!\sim \!0.265 \sqrt{
M C_3^2/2m}$, and $\zeta\!\sim \!5\AA$.
Fig.'s~\ref{fig:coupling-constant-plot} and~\ref{fig:lambda-strength-coupling-plot} show the shifts for various parameter values, the principle results of this paper.
For visualization, we rounded out the resonances, where the dilute gas assumptions fail, by replacing $f \rightarrow \left[1/f + i \alpha\right]^{-1} $ with $\alpha\sim 1$.

Fig.~\ref{fig:coupling-constant-plot} shows the scattering amplitudes for the 1-1 and 1-2 scattering as a function of the mediated interaction strength, with all other parameters fixed at their typical values.  The two vertical lines correspond to zero mediated potential and typical mediated potential strengths.  Fig.~\ref{fig:lambda-strength-coupling-plot} shows the spectral shift as a function of the potential depth and the characteristic length $\lambda$.  The red circle indicates typical parameter values, and the interior of the red box factor of 2 variations of these.
Fig.~\ref{fig:lambda-strength-coupling-plot} shows that for typical values, the shifts are reduced by a factor of 7; simultaneously increasing potential depth and $\lambda$ by a factor of 2 larger than typical values (keeping other parameters typical)  gives a 30-fold decrease.
Similar reductions are possible when other ``typical'' parameters are individually varied (while holding all others constant): 2-fold variations in many of them can produce additional 2-fold reductions. Keeping in mind that the parameters themselves each could believably vary by factors of up to perhaps five,  our calculation demonstrates that the helium-mediated interaction is a consistent and plausible source for the frequency shift reduction. The agreement with experiment is even better when one considers that a 3d control experiment finds an ill-understood factor of 7 discrepancy in the 3d gas~\cite{fn:vas-pc}. Finally, one should note that $\sim\!40\%$ of the change comes from simply using the more accurate formula of Eq.~(\ref{eq:coherent-clock-shift-fundamental-3d}) in place of the approximate Eq.~(\ref{eq:coherent-clock-shift-fundamental-2d}).

\textit{$^3$He film.}---Our theory also accounts for unexpected
effects of adding $^3$He to the film.  It is known that adding $^3$He reduces the
hydrogen surface adsorption energy so
one would naively expect an increase in the confinement length, decrease in density, and decrease in spectral shift.  Instead, adding $^3$He  to the superfluid helium film leads to an observed \textit{increase} in the   shift.

Upon incorporating our theory of the helium surface, changing the adsorption energy $E_a$ has two effects: it increases the confinement length $l$ by $\sim \!2\!-\!10\%$ and increases $\zeta$, the distance of the gas to the helium surface.
We will assume that the fractional change in $\zeta$ is comparable to the fractional change in $l$.
Then Eq.~(\ref{eq:V0-defn}) reveals that
 the mediated potential
 $V_{\text{med}}$ is proportional to $1/\zeta^8\propto 1/l^8$, coming from $\delta$, plus significantly weaker dependence from  $\bar V_\xi$'s dependence on $\lambda$ through $\xi={\lambda/\zeta}$.

Thus, a   $2\!-\!10\%$ increase of $l$ yields a $\sim \!8\!-\!30\%$ decrease of the mediated potential strength.  That such a decrease in mediated potential strength yields a significantly larger spectral shift is clear from Fig.~\ref{fig:coupling-constant-plot}: the $^3$He doping shifts the typical parameters from the left dashed line to the right by $8\!-\!30\%$.  The $1/l$ factor in the spectral shift, due to the decreased 3d density, is in general insufficient to compensate. Using
typical values for parameters yields a range of a $2$-$20\%$ increase of frequency shifts upon adding $^3$He.  This is consistent with the experimentally observed $25\%$ increase, further evidence that the helium surface is playing a crucial role in the experiments.

\textit{Validity of Approximations.}---We have made a number of approximations.  Here we enumerate the most important ones, and discuss which of them need to be addressed in future works through a more sophisticated theory.

The most severe approximation we made is to follow Ref.~\cite{wilson:surface_induced_interaction}, and neglect retardation in the induced potential.  Such effects are relevant when
the phase velocity of the hydrogen excitations $\omega/k$ becomes large compared to $C_3$ at characteristic velocities $\sqrt{k_B T/m}$ and energies $T$~\cite{fn:hazz-unpub}.  For  $T\!\sim \!50 $~mK as in the experiments,
$\omega/k\sim 5$~m/s.  In comparison, typical third sound velocities are $C_3\sim 1$~m/s, so we expect retardation corrections may be significant.  Including the frequency dependence in calculating the spectral shift is challenging, requiring solution of a coupled set of 2d partial differential equations for each $\omega$.  
We expect that
a sharp spectral peak
survives, but with reduced spectral weight.

Additionally, given the long length-scale of the mediated potential, we should also critically examine the assumption that only the long-wavelength limit of the s-wave phase shift is needed to evaluate the cold collision frequency shift.  Eq.~(\ref{eq:coherent-clock-shift-fundamental-3d}) requires that
 the areal interparticle distance ($n_{2D}^{-1/2}$) is larger than the effective range of the potential $R_e$. Since $R_e \sim \!250 \AA$ and $n_{2D}^{-1/2} \sim \!100 \!-\!300\AA$, the approximation with only the  $k\!=\!0$ $s$-wave scattering shift may be sufficiently accurate.
Similarly, the thermal wavelength is $\lambda_T\sim \!30\AA$ while the cold collision regime strictly requires $\lambda_T \gg R_e$.  This can smear out spectral lines somewhat, although since $\lambda_T\!\sim\!R_e$ and the potential is rather shallow, the spectral lines may remain quite sharp.
Here retardation helps us, as the
 slower moving atoms are the dominant contributors to the spectral peak, and these have a substantially larger de Broglie wavelength  -- a factor of 5 for the fastest contributing atoms~\cite{fn:cutoff}

Finally, we have neglected mass renormalization coming from
virtual desorption-adsorption processes.  This approximation is well-justified  since the desorption process is relatively slow~\cite{walraven:film-scattering,statt:thermal-accom},
 implying that the mass is renormalized by at most  a few percent~\cite{eijnde:surface-relaxation}.

\textit{Conclusions.}---In summary, we have shown that incorporating the helium surface into the theory of the hydrogen gas can resolve the $\sim\! 100$-fold theoretical overestimate of the spectral shifts compared to experiment, for a large parameter range consistent with experimental estimates of the parameters.  In addition, we have shown that adding $^3$He to the helium film increases spectral shifts, in agreement with experiment and in contrast to the naive theory.

We have found that the helium surface induces an {\em attractive} interaction between the hydrogen atoms.  It is useful to speculate on what other physical effects this interaction can cause.  For example, can it drive a mechanical instability?  While thermal pressure will stabilize the gas at the temperatures currently being studied, we believe that as the temperature is lowered  that a ``collapse" might occur, similar to the ones seen in atomic gases \cite{randycollapse,muellerbaym,rb85}.  Importantly, the Kosterlitz-Thouless transition requires repulsive interactions, so the mediated interaction may make  eliminate the possibility of achieving superfluidity without significantly altering experimental parameters.

Our calculation suggests an experiment which would further constrain the modeling of this system. Replacing hydrogen with deuterium or tritium drastically changes the parameters in our theory by increasing the binding energy to the helium film, by factors of roughly two and three, respectively~\cite{guyer:polaron}.  However, the form of the theory remains the same, so one can compare the expectations on the basis of our theory with experiment.

Finally, we point out possible ramifications of our theory to ultracold atomic gases and elsewhere. The key to our findings is that in quasi-2d, infinitesimal  attractive interactions generate scattering resonances.  These are pushed to finite depth by  the 3d interaction, but
a genuine 2d interaction easily overwhelms the effects of the 3d interaction.  One can imagine our present theory applying to quasi-2d fermi-bose or bose-bose mixtures where one
system
mediates an interaction for the other species. Interestingly, similar physics might arise in layered systems: one layer effectively mediates an interaction in adjacent layers -- incorporating these effects in a consistent manner could lead to dramatically modified interaction properties for each layer.

\textit{Acknowledgments.}---KH thanks Sergei Vasiliev and Dave Lee for introducing him to the relevant hydrogen experiments, and illuminating discussions.  KH thanks Neil Ashcroft, Sourish Basu, Stefan Baur, Dan Goldbaum, Jarno J\"arvinen, Joern Kupferschmidt, and Sophie Rittner for conversations.  This material is based upon work 
supported by the National Science Foundation through grant No. PHY-0758104.


\begin{thebibliography}{10}
\bibitem{fn:chem-pot}
The ``$\mu_2 - \mu_1$'' comes from the canonical transform introducing chemical potentials into Eq.~(\ref{eq:hydrogen-ham}), and ensures that the absorption spectrum will not depend on $\mu_2$ when initially no particles are in state $|2\rangle$.
\bibitem{fn:cutoff}
Assuming $\omega/k\sim C_3$ is the cutoff, as previously discussed.
\end{thebibliography}
\end{document}